# Calibration setup for ultralow-current transresistance amplifiers

Ilaria Finardi [*,†] and Luca Callegaro [*]
[*]INRIM - Istituto Nazionale di Ricerca Metrologica, Torino, Italy, Email: i.finardi@inrim.it
[†]Politecnico di Torino, Torino, Italy

*Abstract*—We describe a calibration setup for the transresistance of low-current amplifiers, based on the capacitance-charging method. The calibration can be performed in the current range of typical interest for electron counting experiments. The setup implementation is simple and rugged, and is suitable to be embedded in larger experiments where the amplifier is employed. The calibration is traceable to units of capacitance and of time. The base relative accuracy of the implementation is in the $10^{-5}$ range.

## I. INTRODUCTION

In the future revision of the International System of Units (SI) [1] the unit of electrical current, the ampere, will be redefined in terms of the elementary charge $e$. The value of $e$, presently $1.602\,176\,620\,8(98) \times 10^{-9}$ C [2], will be defined as exact (that is, with zero uncertainty). Electron counting experiments [3]–[5] will allow the practical realization of the ampere [6].

The currents generated in these experiments are typically below 1 nA range, and thus must be amplified by large factors ($10^3$ to $10^5$) to be exploited for metrology purposes. The amplification can be performed either with large-ratio cryogenic current comparators [7], [8], or with dedicated current amplifiers [9], [10]. During the development of the experiments, ultralow-current transresistance amplifiers [11]–[13] are commonly employed. The typical transresistance gain $R$ ranges from G$\Omega$ to several T$\Omega$. A traceable gain calibration of these amplifiers allows to verify the fundamental relation $I = nef$, where $I$ is the current generated from single electron devices, $f$ the electron counting frequency, and $n$ is the integer number of electrons counted for each cycle ($n$ ranges from 1 to 3).

In the following, we describe a calibration setup for the transresistance gain of ultralow-current amplifiers based on the capacitance-charging method [14]–[21]. The method allows to produce accurate dc currents in the range 100 fA to 100 pA and is insensitive to non-idealities of the input stage of current amplifiers, such as voltage burden and finite input resistance. Currents can be generated with typical relative uncertainties in the $10^{-5}$ range [14]–[16]; the performances of different measurement setups based on the method have been verified in an international intercomparison [22].

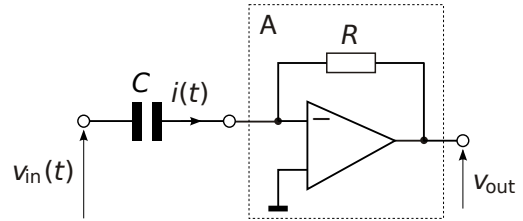

Fig. 1. Principle schematic diagram of the calibration setup, as described in Sec. II. The current $i(t)$ is generated from voltage $v_\text{in}(t)$ by the injection capacitor $C$; the amplifier A, with transresistance $R$ (here ideally associated to its feedback resistance) generates the output voltage $v_\text{out}(t)$.

The setup here proposed allows to perform a calibration of the transresistance gain $R$ of a current amplifier traceable to the capacitance of a gas-dielectric capacitor $C$, and the period $T$ of a low-frequency timebase. The setup configuration is simple and compact, and it is possible to embed the entire setup within the main electron counting experiments, and thus achieve quasi-in-line calibrations of $R$.

An example of calibration of a specific amplifier model (FEMTO mod. DDPCA-300), popular in electron counting experimental setups [11]–[13] and other nanophysics experiments [23]–[25], is given. For this amplifier, the setup allows to calibrate the nominal transresistance gain of 100 G$\Omega$ with a relative uncertainty of about $30 \times 10^{-6}$.

## II. PRINCIPLE OF OPERATION

The operating principle of the calibration setup is shown in Fig. 1. Voltage $v_\text{in}(t)$ is applied to a differentiating capacitor to generate the test current $i(t)$

$$i(t) = C \frac{dv_\text{in}(t)}{dt} \qquad (1)$$

The amplifier A, of which the transresistance gain $R$ has to be calibrated, generates an output voltage $v_\text{out}(t) = R i(t)$, hence the relation

$$R^{-1} = C \, \frac{1}{v_\text{out}(t)} \frac{dv_\text{in}(t)}{dt} \qquad (2)$$

holds. Equation (2) shows that the traceability of the measurement of $R$ is given by $C$, a timebase, and a voltage *ratio*; therefore, absolute voltage traceability is not required.

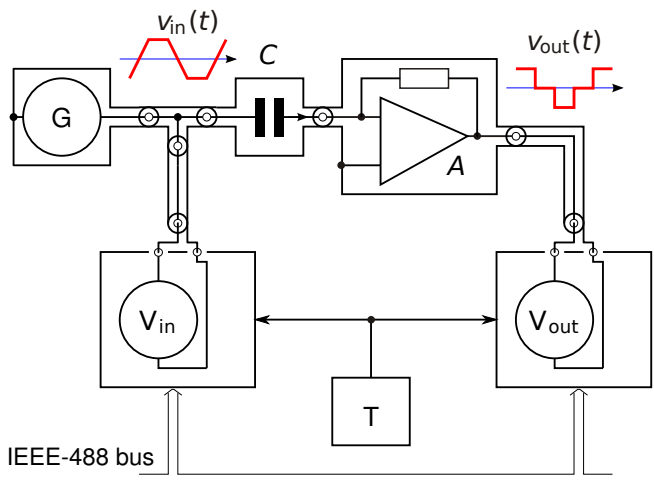

Fig. 2. Schematic diagram of the calibration setup, see Sec. III-A for a description. A pictorial representation of the waveforms of $v_{in}(t)$ and $v_{out}(t)$ is also shown.

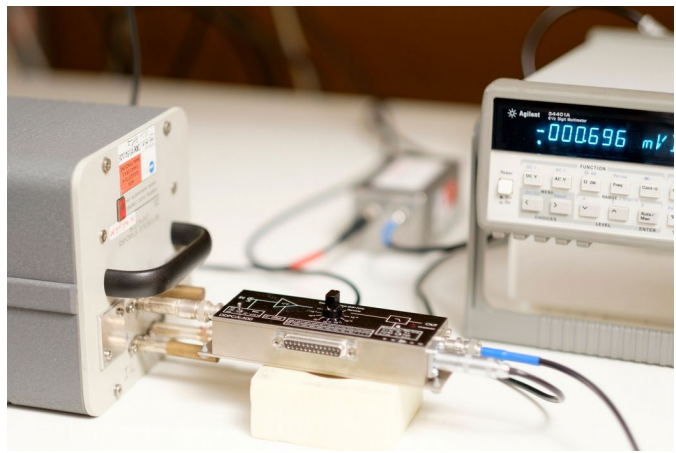

Fig. 4. Detail of setup, showing the direct connection (no cable) of the injection capacitor $C$ (on the left) to the transresistance amplifier A (center of the picture) to minimize currents related to the dielectric absorption in the connection insulators.

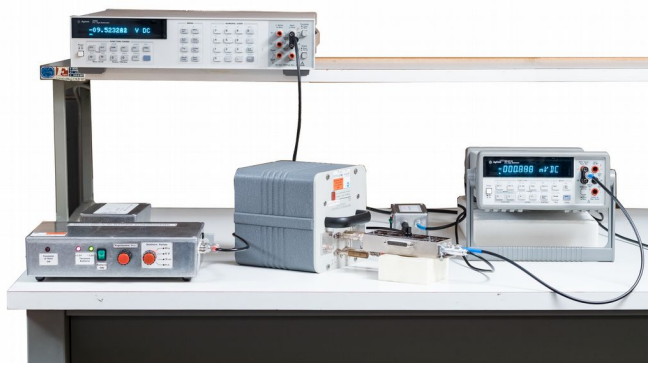

Fig. 3. A photo of the calibration setup corresponding to the schematic diagram of Fig. 2. G is on the bottom left; $C$ and A in the center; $V_{in}$ on top left; $V_{out}$ on the right. A detail of $C$ and A is given in Fig. 4.

The method here proposed is a derivation of a method developed for the calibration of low-current meters [16], [21], where a proper traceability of the $v_{in}$ measurement was however necessary.

## III. IMPLEMENTATION

### A. Setup

The schematic diagram of the calibration setup is shown in Fig. 2, and photo of the same is given in Fig. 3. The source G generates the voltage $v_{in}(t)$, which is applied to the capacitor $C$. The capacitor generates the test current $i(t)$, in agreement with Eq. (1). This test current is injected into the transresistance amplifier A to be calibrated. Both the input voltage $v_{in}(t)$ and the output voltage $v_{out}(t)$ are sampled at regular intervals by the voltmeters $V_{in}$ and $V_{out}$, synchronized since they share the same trigger signal T. The samples of $v_{in}(t)$ and $v_{out}(t)$ are acquired with an interface bus (IEEE-488) for off-line processing.

As shown in Fig. 3, the whole circuit is wired by coaxial leads. To reduce possible effects of dielectric absorption, $C$ and A are connected directly, without any cable, as can be seen in Fig. 4.

The waveform shape of $v_{in}(t)$ generated by G has a symmetric trapezoidal shape, with a very long period, as can be seen in Fig. 5. This specific shape has three different slopes: positive, negative and zero; these slopes correspond to three different nominal calibration current values $+I_{nom}$, $-I_{nom}$ and $I = \pm 0$. The test current $I = \pm 0$ allows to determine the offset of A in the course of the measurement.

### B. Instruments employed

*1) Ramp generator:* G is a purposely-built voltage source. The generated signal $v_{in}(t)$ has a maximum span of $\pm 10$ V, and the ramp sections of the trapezoidal waveform have a slope of $\approx \pm 0.1$ V s$^{-1}$ (adjustable). The positive and negative voltage ramp phases have a duration of $\approx 200$ s each; the phases of constant voltage also have a duration of $\approx 200$ s. Hence, the total period of one $v_{in}(t)$ cycle is $\approx 800$ s. The source is based on analog electronics; it is battery-powered and free-running (thus requiring no control signal), in order to achieve complete galvanic isolation and help to reduce the interferences in the calibration circuit. The source output is generated by an analog pure integrator, which is driven by a three-state (positive, zero, negative) constant current source of adjustable amplitude. The loss in the integrating dielectric capacitor are compensated with an active feedback network, which is manually adjusted in order to achieve the maximum linearity of the voltage ramps. A more complete

TABLE I
INJECTION CAPACITORS EMPLOYED IN THE CALIBRATION SETUP,
LISTED BY NOMINAL CAPACITANCE $C_{\text{nom}}$.

| $C_{\text{nom}}$ | Model |
| --- | --- |
| 1 pF | General Radio mod. 1403-K |
| 10 pF | Sullivan mod. C80001 |
| 100 pF | Sullivan mod. C80002 |
| 1000 pF | General Radio mod. 1404-A [26] |

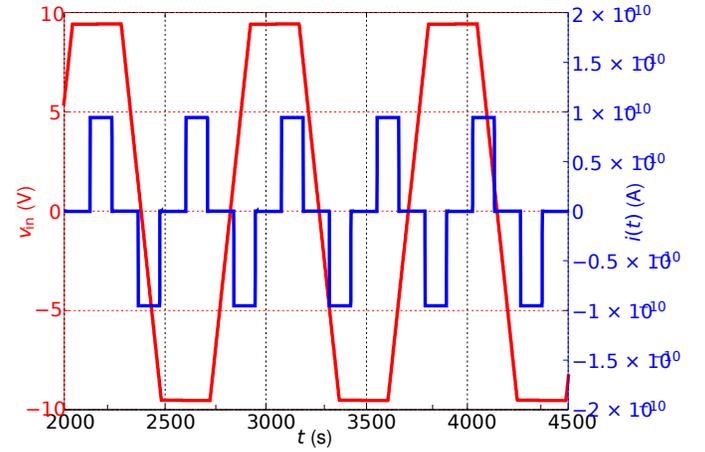

Fig. 5. The outcome of a typical measurement ($R_{\text{nom}} = 10\,\text{G}\Omega$, $C_{\text{nom}} = 1\,\text{nF}$, $I_{\text{nom}} = 95\,\text{pA}$). Red line (—) is the trapezoidal ramp signal $v_{\text{in}}(t)$; blue line (—) is the test current $i(t)$. The sign of $i(t)$ is determined by the sign of the slope of $v_{\text{in}}(t)$; when $v_{\text{in}}(t)$ is constant, $i(t) = 0$.

description of the source is given in Ref. [16].

 2) *Injection capacitor:* $C$ has to be a gas-dielectric (or vacuum) capacitance standard, because all solid-dielectric capacitors show the phenomenon of dielectric absorption [27], which give deviations from Eq. 1. For the current range investigated, commercial standard capacitors having nominal values $C_{\text{nom}}$ in the range 1 pF to 1000 pF are adequate. The specific models employed are listed in Table I. The capacitors have been modified to employ low-dielectric-absorption connectors (Teflon insulation); for the same reason, the solid-dielectric trimming capacitors have been removed.

The value of $C$ is measured as a two terminal-pair standard [28, Ch. 2] with a commercial capacitance bridge (Andeen-Hagerling mod. 2500A) at the frequency of 1 kHz. The calibration is traceable to the Italian national standard of electrical capacitance.

 3) *Voltmeters:* $V_{\text{in}}$ is an Agilent mod. 3458A multimeter, which acquisition is in dc sampling mode, with the autozero and autorange functions disabled. $V_{\text{out}}$ is an Agilent mod. 34401A multimeter, also configured for dc sampling. Both these voltmeters are in external trigger mode, and are synchronously triggered by a precision timer T, at the sampling frequency of $\approx 950\,\text{mHz}$. All samples are acquired via the IEEE-488 bus and off-line processed. Although not required by the proposed method (see discussion in Sec. II), the voltmeters are routinely calibrated, with traceability to the Italian national standard of dc voltage.

### C. The device under test

The calibration setup has been tested with a FEMTO mod. DDPCA-300 transresistance amplifier A. The amplifier has a nominal transresistance range $R_{\text{nom}}$ manually switchable from 10 kΩ to 10 TΩ and is specified to be stable for capacitance at the input up to 10 nF, therefore for all capacitance standards of Tab. I. The voltage output span is ±10 V; the current noise is dependent on $R_{\text{nom}}$ and reaches 200 aA Hz$^{-\frac{1}{2}}$ in the highest gain ranges. The specified accuracy of $R_{\text{nom}}$ is ±1% with a temperature coefficient of $3 \times 10^4\,\text{K}^{-1}$. The amplifier has a configurable output lowpass filter; all measurements reported have been performed in the so-called *full bandwidth* (dc to 400 Hz) mode.

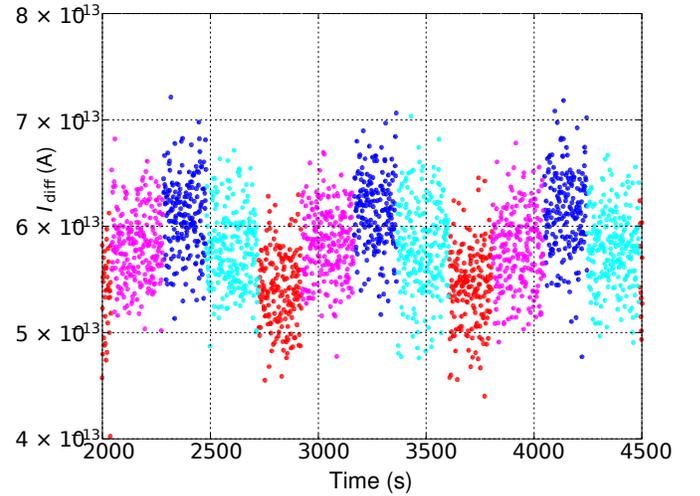

Fig. 6. Time sequence of the amplifier equivalent error the input $\Delta i(t)$ (see Sec. IV for the definition). The four different dot colors correspond to the four different phases of $v_{\text{in}}(t)$. • corresponds to $v_{\text{in}}$ positive ramp slope, and to $i(t) = +I_{\text{nom}}$. • negative ramp slope, $i(t) = -I_{\text{nom}}$. • $v_{\text{in}}(t)$ constant positive, $i(t) = +0$. • $v_{\text{in}}(t)$ constant negative, $i(t) = -0$. The offset of A is computed from the average of the $i(t) = +0$ and $i(t) = -0$ phases.

### IV. RESULTS

#### A. Description of the measurement

The setup has been employed to calibrate the transresistance nominal settings $R_{\text{nom}} = 10\,\text{G}\Omega$, 100 GΩ, 1 TΩ and 10 TΩ of A. All measurements have been performed in a shielded and thermostated (23.0(5) C) room. Each calibration has been achieved by running the system for about 50 cycles of $v_{\text{in}}(t)$, corresponding to a total measurement time of 10 h. The calibration strategy and the related acquisiton, data processing software and uncertainty analysis described in [16], [22] apply.

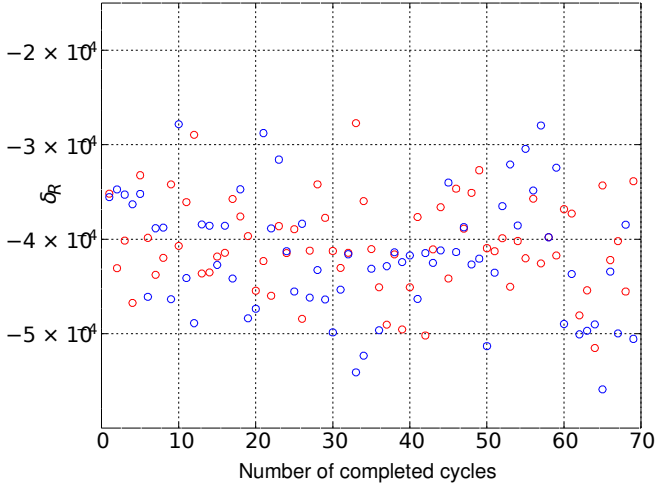

Fig. 7. The amplifier transresistance gain error $\delta R = (R - R_{nom})/R_{nom}$ versus measurement time, evaluated for each positive (○) and negative (○) semicycles of $v_{in}(t)$.

TABLE II
RESULTS OF THE TRANSRESISTANCE AMPLIFIER CALIBRATION

| $R_{nom}$ | $I_{nom}$ | $C_{nom}$ | $\delta R$ |
|---|---|---|---|
| 10 GΩ | 95 pA<br>−95 pA | 1 nF | $-4.06(34) \times 10^4$<br>$-4.18(36) \times 10^4$ |
| 100 GΩ | 9.5 pA<br>−9.5 pA | 100 pF | $-3.06(4) \times 10^3$<br>$-3.11(4) \times 10^3$ |
| 1 TΩ | 9.5 pA<br>−9.5 pA | 100 pF | $-2.90(3) \times 10^3$<br>$-4.32(3) \times 10^3$ |
| 1 TΩ | 0.95 pA<br>−0.95 pA | 10 pF | $-3.29(9) \times 10^3$<br>$-3.14(7) \times 10^3$ |
| 10 TΩ | 0.95 pA<br>−0.95 pA | 10 pF | $-5.77(8) \times 10^3$<br>$-5.76(6) \times 10^3$ |
| 10 TΩ | 0.095 pA<br>−0.095 pA | 1 pF | $-5.83(18) \times 10^3$<br>$-6.03(2) \times 10^3$ |

To avoid possible systematic error in the calibration caused by noise clipping, the calibration nominal currents $\pm I_{nom}$ are chosen to be slightly lower (in absolute value) than the corresponding decadic value. In particular, the $I_{nom}$ values chosen are: $\pm 0.95$ pA, $\pm 9.5$ pA, and $\pm 95$ pA.

The measurement example of Fig 5, Fig. 6 and 7 refer to the following calibration conditions: $R_{nom} = 10$ GΩ, $C_{nom} = 1$ nF, $I_{nom} = \pm 95$ pA.

Fig. 5 shows the time sequence of the samples of $v_{in}(t)$ (measured by $V_{in}$) and of $i(t)$ as determined by Eq. (1) over a few measurement cycles.

Fig. 6 displays the amplifier equivalent error at the input $\Delta i(t) = R_{nom}^{-1} v_{out}(t) - i(t)$, here defined as the deviation of the current reading $R_{nom}^{-1} v_{out}(t)$ (computed from the amplifier voltage output $v_{out}(t)$ with the nominal transresistance $R_{nom}$), and the calibration current $i(t)$.

Fig. 7 shows the transresistance gain error $\delta R = (R - R_{nom})/R_{nom}$ evaluated for each positive and negative semicycles of $v_{in}(t)$.

The outcome of a typical calibration result can be seen in Fig. 7: for each positive and negative ramp in a cycle the figure reports the corresponding relative deviation $\delta R = (R - R_{nom})/R_{nom}$ of the transresistance $R$ to be calibrated from the nominal value $R_{nom}$. The values reported in Tab. II correspond to the average of the $\delta R$ point values over the whole measurement cycles.

### B. Calibration summary

The calibrated values of $\delta R$ for different test currents and nominal transresistance of the amplifier are reported in Tab. II together with the corresponding uncertainty (see Sec. IV-C). For the nominal transresistances $R_{nom} = 1$ TΩ and 10 TΩ the calibration is performed with two different test currents. For $R_{nom} = 1$ TΩ significant differences among the $\delta R$ values obtained with different current magnitudes and sign; these discrepancies deserve further investigation.

### C. Traceability and uncertainty

As explained in Sec. II, the traceability of the calibration is provided by the calibrated value of the injection capacitor $C$ and the period of timebase T. Several other effects contribute the measurement uncertainty. Among them we list the measurement noise, possible frequency dependencies of $C$ (which is measured at 1 kHz but employed at mHz frequency in the setup), current leakages in $C$, tracking and nonlinearities of $V_{in}$ and $V_{out}$. These influence quantities are under investigation and therefore the uncertainties reported in Tab. II should be considered preliminary and not including the in-use uncertainty (that will account for time and environmental drifts).

## V. CONCLUSIONS AND OUTLOOK

The proposed setup can calibrate the transresistance gain of amplifier suitable for the measurement of ultralow-valued dc currents. The uncertainty achieved is one or two order of magnitudes better than the corresponding manufacturer specifications. The calibration uncertainties reported is still preliminary, however the calibration accuracy is one-two orders of magnitude better than typical manufacturer accuracy specifications. A complete uncertainty budget will be reported at the Conference.

The method is simple and can be embedded in an electron-counting experiment with relative ease. As Eq. 2 shows, the method requires traceability to capacitance and time units, since it involves only voltage ratios: this opens the possibility of further simplification in the measurement setup, by performing voltage ratio measurements with a single, two-channel sampling instrument which would not require absolute voltage calibration. This alternative setup is presently under investigation, and will be reported at the Conference.

## VI. Acknowledgments

The authors are indebted with Vincenzo D'Elia, INRIM, for help in the realization of the measurement setup; and with Luca Croin, INRIM, for the photographic material.